\documentclass[prd,superscriptaddress,showpacs,nofootinbib,floatfix,eqsecnum,
groupedaddress,superscriptaddress
,eqsecnum,twocolumn
]{revtex4}

\usepackage{amssymb,amsmath}
\usepackage[]{graphicx}
\usepackage{amsfonts,bm,latexsym,psfrag,multirow}

\usepackage{dcolumn,epsfig}

\newcommand{\hL}{\hat{\mathbf{L}}}
\newcommand{\hS}{\hat{\mathbf{S}}}

\newcommand{\vz}{\mathbf{z}}

\newcommand{\vL}{\mathbf{L}}
\newcommand{\vn}{\mathbf{n}}

\newcommand{\beq}{\begin{equation}}
\newcommand{\eeq}{\end{equation}}
\newcommand{\bes}{\begin{subequations}}
\newcommand{\ees}{\end{subequations}}
\newcommand{\bea}{\begin{eqnarray}}
\newcommand{\eea}{\end{eqnarray}}
\newcommand{\ba}{\begin{array}}
\newcommand{\ea}{\end{array}}
\newcommand{\beqn}{\begin{eqnarray*}}
\newcommand{\eeqn}{\end{eqnarray*}}

\newcommand{\f}[2]{\frac{#1}{#2}}
\newcommand{\lisa}{{\em LISA}~}

\newcommand{\I}{\mbox{\rm i}} 

\def\jnl@style{\it}
\def\aaref@jnl#1{{\jnl@style#1}}

\def\aap{\aaref@jnl{A\&A}}                
	\def\aapr{\aaref@jnl{A\&A~Rev.}}          
\def\aaps{\aaref@jnl{A\&AS}}              
\def\mnras{\aaref@jnl{MNRAS}}             

\def\bt{\bar \theta}
\def\bph{\bar \phi}

\def\nn{\nonumber}

\newlength{\sizeonefig}
\newlength{\sizetwofig}
\begin{document}

\title[]{Bounding the mass of the graviton with gravitational waves: Effect of spin precessions in massive black hole binaries}

\author{Adamantios Stavridis}\email{astavrid@physics.wustl.edu}
\affiliation{McDonnell Center for the Space Sciences, Department of
Physics, Washington University, St.\ Louis, Missouri 63130}

\author{Clifford M. Will}\email{cmw@wuphys.wustl.edu}
\affiliation{McDonnell Center for the Space Sciences, Department of
Physics, Washington University, St.\ Louis, Missouri 63130}
\affiliation{GReCO, Institut d'Astrophysique de Paris, CNRS,\\ 
Universit\'e Pierre et Marie Curie, 98 bis Bd.\ Arago, 75014 Paris, France}

\begin{abstract}
Observations of gravitational waves from massive binary black hole systems at cosmological distances 
can be used to search for a dependence of the speed of propagation of the waves on wavelength, and thereby to bound the mass of a hypothetical graviton.   We study the effects of precessions of the spins of the black holes and of the orbital angular momentum on the process of parameter estimation based on the method of matched filtering of gravitational-wave signals vs.\ theoretical template waveforms.   For the proposed space interferometer {\em LISA}, we show that precessions, and the accompanying modulations of the gravitational waveforms, are effective in breaking degeneracies among the parameters being estimated, and effectively restore the achievable graviton-mass bounds to levels obtainable from binary inspirals {\em without} spin.
For spinning, precessing binary black hole systems of equal masses $10^6 \,M_\odot$ at 3 Gpc, the lower bounds on the graviton Compton wavelength achievable are of the order of $5 \times 10^{16}$ km.
\end{abstract}

\vspace{0.5cm}


\maketitle

\section{Introduction and Summary}
\label{Intro}
The anticipated launch of the Laser Interferometer Space Antenna ({\em LISA}) in the 2020 time frame will provide
a promising new tool for doing astrophysics with massive binary black hole systems.
The inspiral and merger of massive black holes (MBH) with masses  of the order of $10^5$ - $10^7$ $M_{\odot}$
will be detectable to large distances in {\em LISA}'s sensitive
frequency band   between $10^{-5}$
and $1$ Hz. 
The detection of gravitational waves (GW) from MBH systems will allow us to infer important astrophysical and
astronomical information, such as the masses and spins of the black holes, the location of the system on the sky
and its distance from the solar system.  

Another important aspect of MBH binaries
is the possibility of testing general relativity itself.  In previous papers we have studied the bounds that could be placed on alternative theories of gravity such as scalar-tensor theories of the Brans-Dicke type, and theories in which gravitational waves propagate with a wavelength-dependent speed, as if the ``graviton'' were massive~\cite{
1994PhRvD..50.6058W,
1998PhRvD..57.2061W,
2002PhRvD..65d2002S,
2004CQGra..21.4367W,
2005PhRvD..71h4025B,
Arun:2009pq}.  Specifically, in~\cite{2005PhRvD..71h4025B} we showed that the inclusion of aligned, non-precessing spins weakens the bounds obtainable on the graviton mass by almost an order of magnitude.  This is because the parameters that characterize the inpiraling binary are highly correlated, so that the addition of parameters (the spins) into the estimation process effectively dilutes the available information, leading to weakened bounds or estimates on most parameters.

However, Vecchio \cite{2004PhRvD..70d2001V} pointed out that when the effects of precession of spins are incorporated 
into the gravitational waveforms, i.e.\ when the spins are not aligned with the orbital angular momentum, the accuracy of parameter estimation can be improved.   He studied
the so called ``simple precession'' case where either one of the bodies has zero spin,  
or the black hole masses are equal, and only spin-orbit interactions are included.  The modulations of the amplitude and phase of the gravitational waveform induced by the precession of the spin(s) and by the precession of the orbital plane effectively {\em adds} information to the estimation process, partially decouples some of the parameters, and thus leads to restored accuracy.  Lang and Hughes~\cite{2006PhRvD..74l2001L} extended Vecchio's work to include arbitrary spins and masses, and also spin-spin interactions, and found significant
improvements in the accuracy of mass measurements as well as sky localization. In addition, they showed that the magnitudes of the spins of the binary members, especially for low redshift 
systems at $z \simeq 1$, could be measured with accuracies of the order of $10^{-2}$. 

In this paper we describe the results of an independent code written by one of us (AS) for analysing binary inspiral with precessing spin and for carrying out parameter estimation based on the method of matched filtering, but extended  
to include the effects of a massive graviton.   In addition to confirming the central conclusions of Lang and Hughes~\cite{2006PhRvD..74l2001L}, we show that spin precessions significantly improve the bounds that can be placed on the mass of the graviton.   In parallel work, we have shown that 
including higher signal harmonics in the PN waveform (but without spins) also leads to improved bounds on the graviton mass~\cite{Arun:2009pq}.

\begin{figure}[!t]
\begin{center}
\includegraphics[width=2.8in,angle=-90]{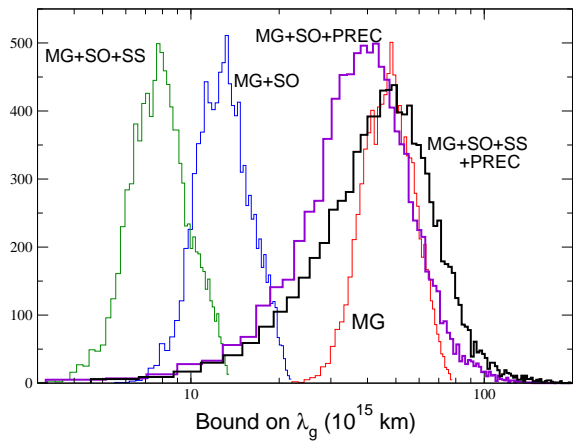}
\caption{Distribution of lower bounds on the graviton Compton wavelength 
$\lambda_{g}$ (in units of $10^{15}$ km) for $10^4$ equal-mass ($10^6 \, M_{\odot}$) black-hole binaries at redshift $z= 0.55$, or a luminosity distance 3 Gpc, randomly located on the sky.  Number of bins is set to 50.  First three histograms (narrow lines; red, blue, green in the color version) assume either no spins or  aligned spins with spin-orbit (SO) and/or spin-spin (SS) coupling.  Final two histograms (thick lines; violet and black in the color version) include precessions induced by non-aligned spins}
\label{fig:BBW_all}
\end{center}
\end{figure}

\begin{figure}[!t]
\psfragscanon
\psfrag{xlabel}{$\lambda_{\rm g}$ ($10^{15}$ km) }
\begin{center}
\includegraphics[width=2.8in,angle=-90]{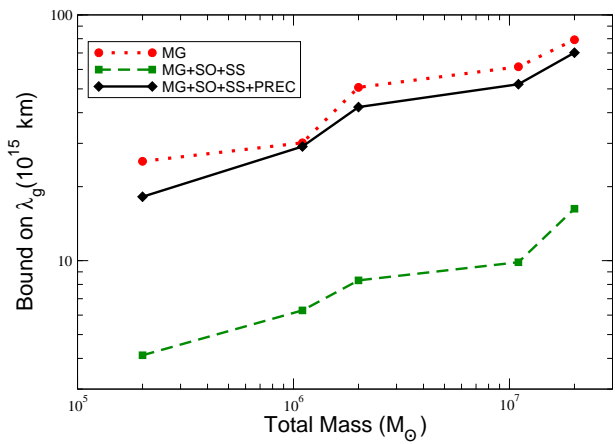}
\caption{Median lower bounds on the graviton Compton wavelength 
$\lambda_{g}$ (in units of $10^{15}$ km) for $10^4$  black-hole binaries at redshift $z= 0.55$, or a luminosity distance 3 Gpc, randomly located on the sky.  Systems contain black holes of mass $(1,1) \times 10^5$, $(1,10) \times 10^5$, $(1,1) \times 10^6$, $(1,10) \times 10^6$ and $(1,1) \times 10^7 \, M_\odot$, from left to right, respectively.}
\label{fig:boundvsMass}
\end{center}
\end{figure}

Our main conclusion, shown in Figs.\ \ref{fig:BBW_all} and \ref{fig:boundvsMass} is that the inclusion of spin precession effects
increases the lower bound on the graviton Compton wavelength $\lambda_{\rm g}$ by almost an order of magnitude, on average, with
respect to the one calculated for the same non precessing system.
Recall that $\lambda_g$ is related to the mass of the graviton by $\lambda_g = h/m_g c$, where $h$ is Planck's constant and $c$ is the speed of light, so that a lower bound on $\lambda_g$ represents an upper bound on $m_g$. 

Indeed, the new bounds, labeled MG+SO+PREC and MG+SO+SS+PREC in Fig.\ {\ref{fig:BBW_all}, which incorporate spin-orbit (SO) effects only and spin-orbit and spin-spin (SS) effects, respectively, along with the effect of a massive graviton (MG), are comparable to those inferred from an identical system without spin effects at all, labeled MG.   This improvement is independent of mass, as seen in Fig.\  \ref{fig:boundvsMass}, which plots median lower bounds on $\lambda_g$ for systems without spin (MG), with non precessing spins (MG+SO+SS), and with precessing spins 
(MG+SO+SS+PREC) for various pairs of masses spanning two orders of magnitude in total mass.   

The rest of the paper provides the details of the analysis behind our main conclusion. In Sec.\ \ref{Signal} we review the construction of gravitational waveform templates and the orbital dynamics when spin precessions are incorporated.  In Sec.\ \ref{Parameter} we describe the parameter estimation process based on the method of matched filtering, and in Sec.\ \ref{calculations} we present the results.  Section \ref{conclusions} presents concluding remarks.  Throughout the paper we use units in which $c=G=1$. 

\section{Gravitational waveform and orbital dynamics including spin precessions }
\label{Signal}

In this section we give a brief overview of the assumptions made for the GW signal used 
for our calculations.
The waveform emitted by an inspiraling black hole binary system can be described accurately 
by the post-Newtonian approximation developed by several groups (see for example~\cite{1996CQGra..13..575B}; for a review of
the post-Newtonian approximation for gravitational wave emission from 
inspiraling binaries see~\cite{lrr-2006-4}).
For our study we made the following assumptions, some physically justified
and some imposed for simplicity.
(i) We take into account only the inspiral phase of the signal, ignoring the merger and ringdown part.  The bound on the graviton mass will be dominated by information from the inspiral phase where the wavelength of the signal varies over many orders of magnitude.
(ii) We assume the restricted second post-Newtonian (2PN) approximation, in which the amplitude of the signal is evaluated to the lowest, Newtonian order, while the phase is evaluated to 2PN order.  (Ref.\ \cite{Arun:2009pq} goes beyond this approximation, but does not include spins.)
(iii) We use the stationary phase approximation (SPA) for calculating the Fourier transform of the signal.
(iv) We assume that the orbits are quasi-circular.

With these assumptions and following \cite{2006PhRvD..74l2001L}, we express the Fourier transform of the GW signal as
\begin{equation}
\tilde{h}_I(f) = \sqrt{\frac{5}{96}}
\frac{\pi^{-2/3} 
{\cal M}^{5/6}}{D_L} A^{I}[t(f)] \; f^{-7/6} \;
e^{ \I \Phi_I } \, ,
\label{eq:freqdomainsignal}
\end{equation}
where $f$ is the frequency of the wave; 
$\mathcal{M}$ is the ``chirp mass'' of the system given by $\mathcal{M} = \eta^{3/5} M$, with
$M = m_1 + m_2$ and $\eta = m_1 m_2 / M^2$; 
$I=1,2$ labels two possible combinations of data from the three arms of {\em LISA}.  The quantity $t(f)$ is the time at which the emitted gravitational-wave frequency equals $f$.  
The                                                                                                                                                                                                                                                                                                    distance of the source $D_L$, is  given as a function 
of the redshift by the expression,
\begin{equation}
D_L(z) = \frac{(1+z)}{H_0}\int_0^z
\frac{dz^\prime}{\sqrt{\Omega_M(1+z')^3 + \Omega_\Lambda}} \, ,
\label{eq:Dzrelation}
\end{equation}
with the cosmological parameters having the values $\Omega_M=0.25$, $\Omega_{\Lambda}=0.75$, $H_0=75 \, {\rm km \, s^{-1} \, Mpc^{-1} }$, 
following the latest fits by the WMAP mission~\cite{Spergel:2006hy}.

The amplitude of the wave,
$A^{I} [ t(f) ]$, is  given by the expression,
\begin{eqnarray}
A^{I}(t) &=& \frac{\sqrt{3}}{2} \left ( [ 1+(\mathbf{\hat{L}} \cdot
\mathbf{\hat{n}})^2]^ 2F_I^+(t)^2 
\right .
\nonumber
\\
&& \quad \left .+ 4(\mathbf{\hat{L}} \cdot
\mathbf{\hat{n}})^2F_I^{\times}(t)^2 \right )^{1/2} \,,
\label{eq:polamp}
\end{eqnarray}
where, $\mathbf{\hat{L}}$ and $\mathbf{\hat{n}}$ are unit vectors in the directions of
the source orbital angular momentum and the line of sight to the source,
respectively. The \lisa antenna pattern functions for one pair of arms $F_1^{+,\times}$
are given by the expressions \cite{1998PhRvD..57.7089C}, 
\bea
F^+_{1}(\theta_S,\phi_S,\psi_S) &=& \frac{1}{2}(1+\cos^2\theta_S)\cos
2\phi_S \cos 2\psi_S 
\nonumber \\
&& - \cos \theta_S \sin 2\phi_S \sin 2\psi_S \, , \nonumber
\label{eq:FI+}\\
F^{\times}_{1}(\theta_S,\phi_S,\psi_S) &=&
\frac{1}{2}(1+\cos^2\theta_S)\cos 2\phi_S \sin 2\psi_S
\nonumber \\
&&+ \cos \theta_S \sin 2\phi_S \cos 2\psi_S \, ,
\label{eq:FIx}
\eea
where $\theta_S$ and $\phi_S$ are the spherical angles for the
binary's line of sight $\mathbf{\hat{n}}$ in a frame in which the three {\em LISA} spacecraft are at rest, 
and $\psi_S$ is the polarization angle of the wave in the same frame given
by the expression
\begin{equation}
\tan \psi_S = \frac{\mathbf{\hat{q}}\cdot
\mathbf{\hat{z}}}{\mathbf{\hat{p}}\cdot \mathbf{\hat{z}}} =
\frac{\mathbf{\hat{L}}\cdot \mathbf{\hat{z}}-(\mathbf{\hat{L}}\cdot
\mathbf{\hat{n}})(\mathbf{\hat{z}}\cdot
\mathbf{\hat{n}})}{\mathbf{\hat{n}}\cdot (\mathbf{\hat{L}}\times
\mathbf{\hat{z}})} \, .
\label{eq:psiS}
\end{equation}
The unit vector $\mathbf{\hat{z}}$ is orthogonal to the plane of the {\em LISA} satellites, while $\mathbf{\hat p}$ and $\mathbf{\hat q}$ are axes orthogonal to $\mathbf{\hat n}$, defined as
$\mathbf{\hat p} = \mathbf{\hat n} \times \mathbf{\hat L} / | \mathbf{\hat n} \times \mathbf{\hat L} |$
and
$\mathbf{\hat q} = \mathbf{\hat p} \times \mathbf{\hat n}$; 
they are the principal axes of the wave, i.e. defined such that the two polarizations are exactly 
$90^{\rm o}$ out of phase.
For the second ``detector'' (actually a linear combination of outputs from the three \lisa arms such that the noise is independent of  the noise in the two arms that make up detector 1) the expressions are,
\begin{eqnarray}
F^+_{2}(\theta_S,\phi_S,\psi_S) &=&
F^+_{1}(\theta_S,\phi_S-\frac{\pi}{4},\psi_S) \, , \nonumber
\label{eq:FII+}\\
F^{\times}_{2}(\theta_S,\phi_S,\psi_S) &=&
F^{\times}_{1}(\theta_S,\phi_S-\frac{\pi}{4},\psi_S) \, .
\label{eq:FIIx}
\end{eqnarray}

In order to use these expressions for our calculations they must be transformed to a 
coordinate system tied to the ecliptic. Taking into account the ``cartwheel'' motion of the {\em LISA} array as it orbits the Sun, we use expressions in Ref.~\cite{1998PhRvD..57.7089C},
\begin{eqnarray}
\cos \theta_S &=& \frac{1}{2} \cos\bar{\theta}_S  - \frac{\sqrt{3}}{2} 
\sin \bar{\theta}_S \cos[ \bar{\phi}(t) - \bar{\phi}_S ] \,,
\nonumber
\\
\phi_S &=& \alpha_0 + 2\pi \frac{t}{T} 
+ \tan^{-1} \beta  \,, \nonumber
\\
\beta  &=&
\frac{\sqrt{3} \cos\bar{\theta}_S + \sin\bar{\theta}_S \cos[ \bar{\phi}(t) - \bar{\phi}_S ] }
{2 \sin\bar{\theta}_S \sin[ \bar{\phi}(t) - \bar{\phi}_S ] } \,,
\end{eqnarray}
where,  $\bar{\theta}_S$ and $\bar{\phi}_S$ denote the fixed direction to the source, and 
$\bar{\phi}(t)= \bar{\phi}_0 + 2 \pi {t}/{T}$ denotes barycentric longitude of the detector's 
center-of-mass as it orbits the Sun,
where $T$ is one year and $\bar{\phi}_0$, $\alpha_0$ are arbitary
orientation constants usually chosen to be zero. 
The polarization angle $\psi_S$ is written in terms of barycentric angles using Eq. (\ref{eq:psiS}) and the expressions
\cite{2005PhRvD..71h4025B},
\bea
\hat{\vz} \cdot \hat{\vn} &=& \frac{1}{2} \cos\bar{\theta}_S  - \frac{\sqrt{3}}{2} 
\sin \bar{\theta}_S \cos \bigl ( \bar{\phi}(t) - \bar{\phi}_S \bigr ) \,,
\nonumber \\
\hat \vL \cdot \hat \vz &=& \frac{1}{2} {\rm cos}\,\bt_L -
\frac{\sqrt{3}}{2} {\rm sin}\; \bt_L  {\rm cos}\,\bigl(\bph(t) -
\bph_L \bigr)\;,
\nonumber \\
\hat \vL \cdot \hat \vn &=& {\rm cos}\; \bt_L {\rm cos}\;\bt_S 
\nonumber \\
&&+
{\rm sin}\,\bt_L  \; {\rm sin}\,\bt_S \; {\rm cos}\,\bigl(\bph_L -
\bph_S \bigr) \;,
\nonumber \\
\hat \vn \cdot (\hat \vL \times \hat \vz) &=& \frac{1}{2} \,{\rm
sin}\,\bt_L \; {\rm sin}\,\bt_S\;
{\rm sin}\,\bigl(\bph_L - \bph_S \bigr) \nonumber \\
&&
+ \frac{\sqrt{3}}{2} \left [
 {\rm cos}\,\bt_L \; {\rm sin}\,\bt_S \; {\rm sin}\,(\bph(t)-\bph_S) 
 \nonumber \right . \\
&& \left . -
 {\rm cos}\,\bt_S \; {\rm sin}\,\bt_L \; {\rm sin}\,(\bph(t)-\bph_L)
 \right ] \;.
\eea
where $\bt_L$ and $\bph_L$ are the polar and azimuthal angles of the orbital angular momentum vector $\hat \vL$ in barycentric coordinates.

In order to determine $t(f)$ in Eq.\ (\ref{eq:freqdomainsignal}), we use
the rate at which the observed frequency changes because of the emission of gravitational radiation 
by the binary system and because of the propagation delay induced by a massive graviton, as given by the expression \cite{1995PhRvL..74.3515B},
\begin{eqnarray}
\frac{df}{dt} &=& \frac{96}{5 \pi {\cal M}^2}\,(\pi {\cal M} f)^{11/3}\,
\left \{ 1 + \beta_g\,(\pi {\cal M} f)^{2/3}
 \right. \nn \\
& & \left.
 -\left(\frac{743}{336}+\frac{11}{4}\eta\right)(\pi
M)^{2/3} 
+ (4\pi-\beta)(\pi Mf) \right. \nn \\
& & \left. + \left(\frac{34103}{18144} + \frac{13661}{2016}\eta +
\frac{59}{18}\eta^2 + \sigma \right)(\pi Mf)^{4/3}
 \right \}\,, \nn \\
\label{eq:fdot}
\end{eqnarray}
where
\begin{equation}
\beta_g = \frac{\pi^2 D\,{\cal M}}{\lambda_g^2\,(1+z)} \, ,
\label{eq:betag}
\end{equation}
describes the contribution of the massive graviton.  Its effect 
is to alter the time of arrival of the wavefronts 
for a given frequency, as a function of the Compton wavelength $\lambda_g$ and a distance 
parameter defined as \cite{1998PhRvD..57.2061W},
\begin{equation}
D = \frac{1+z}{H_0} \int_0^z \frac{dz'}{(1+z')^2 
\sqrt{\left[\Omega_M(1+z')^3+\Omega_\Lambda\right]}} \,.
\label{Eq:Dgrav}
\end{equation}

The parameter $\beta$, describing spin-orbit interactions, is given by
\begin{equation}
\beta = \frac{1}{12} \sum_{i=1}^2 \chi_i \left [113 \left( \frac{m_i}{M} \right)^2 + 75
\eta \right ] \hL \cdot \hS_i \,,
\label{eq:beta}
\end{equation}
and the parameter $\sigma$ describing spin-spin interactions is
given by
\begin{equation}
\sigma = \frac{\eta}{48} \chi_1 \chi_2 \left[ 721 ( \hL
\cdot \hS_1 ) ( \hL \cdot \hS_2  )  - 247 (\hS_1 \cdot \hS_2)  \right],
\label{Eq:sigma}
\end{equation}
where $\chi_{i} = S_{i} / m_i^2$, is the dimensionless spin parameter for each body.

To get the relation between the time elapsed and the frequency, one
has to integrate Eq.\ (\ref{eq:fdot}). 
When spin precessions are taken into account, both the spin-orbit and spin-spin
coefficients $\beta$ and $\sigma$ are oscillating functions of time  around an average value; however,
as shown in \cite{1994PhRvD..49.2658C}, the amplitude of the oscillations 
is small so one can, without significant loss of acccuracy, assume that 
they are constant for the purpose of the integration.  The result is
\begin{widetext}
\begin{eqnarray}
t(f) &=& t_c -\frac{5}{256}{\mathcal M}(\pi {\mathcal M} f)^{-8/3} 
\left\{ 1 + \frac{4}{3} \beta_g \,(\pi {\cal M} f)^{2/3}
- \frac{4}{3}\left(\frac{743}{336} +
\frac{11}{4}\eta \right)(\pi Mf)^{2/3} - \frac{8}{5}(4\pi - \beta)(\pi Mf) \right. \nn \\
&+& \left.  2\left(\frac{3058673}{1016064} +
\frac{5429}{1008}\eta + \frac{617}{144}\eta^2 - \sigma\right)(\pi
Mf)^{4/3}\right\} \,.
\label{eq:PNtime}
\end{eqnarray}

In our calculations we use the above expression wherever necessary 
to express time as a function of frequency, but we insert the frequency-dependent values of $\beta$ and $\sigma$ that come out of the numerical integration 
of the spin precession equations, as described below.  Although this is a slightly inconsistent procedure, we do not expect it to have a large effect, since the spin-orbit and spin-spin terms are high-order PN corrections, and thus are relatively small.

The phase $\Phi_I$ in Eq. (\ref{eq:freqdomainsignal}) has several terms that describe different effects 
contributing to the phasing of the gravitational wave, in the form
\beq
\Phi_I = \Psi(f)  - \varphi_\mathrm{pol}^{I}[t(f)] - \varphi_D[t(f)] - \delta_p\Phi[t(f)] \, .
\eeq
The first term $\Psi(f)$, is the phasing function at 2PN order arising from the internal dynamics of the binary system, given by the expression
\cite{2005PhRvD..71h4025B},
\begin{eqnarray}
\Psi(f) &=& 2\pi f t_c - \Phi_c - \frac{\pi}{4} + \frac{3}{128}(\pi
{\mathcal M} f)^{-5/3}
\left[1
-\frac{128}{3} \, \beta_g \,(\pi {\cal M} f)^{2/3}
+ \frac{20}{9}\left(\frac{743}{336} +
\frac{11}{4}\eta\right)(\pi Mf)^{2/3}  \right. \nn \\
&-& \left. 4(4\pi - \beta)(\pi Mf) + 10\left(\frac{3058673}{1016064} +
\frac{5429}{1008}\eta
+ \frac{617}{144}\eta^2 - \sigma \right)(\pi
Mf)^{4/3}\right] \, ,
\end{eqnarray}
\end{widetext}
where $t_c$ and $\Phi_c$, are the time and the phase of coalescence respectively.  Here, as in the calculation of $t(f)$, we hold $\beta$ and $\sigma$ fixed during the required integrations, and then insert the time-varying values afterward.

The term $\varphi_\mathrm{pol}^{I}[t(f)]$, often called the ``polarization phase'',  arises from the conversion of the real signal into an amplitude (\ref{eq:polamp}) and a phase, and is   
given by the expression,
\begin{equation}
\varphi_\mathrm{pol}^{I}(t) =
\tan^{-1}\left[\frac{2(\mathbf{\hat{L}}\cdot
\mathbf{\hat{n}})F_I^{\times}(t)}{[1+(\mathbf{\hat{L}}\cdot
\mathbf{\hat{n}})^2]F_I^+(t)}\right] \, .
\label{eq:polphase}
\end{equation}
The term $\varphi_D[t(f)]$, is the ``Doppler phase'', arising 
from the varying arrival time of the signal as the detector moves around the sun, given for both detectors by the expression,
\begin{equation}
\varphi_D(t) = 2\pi f R_{\oplus}
\sin{\bar{\theta}_S}\cos[\bar{\phi}(t)-\bar{\phi}_S] \, ,
\label{eq:dopphase}
\end{equation}
where $R_{\oplus}= 1 {\rm \; AU}$.

Finally, the term $\delta_p {\Phi}[t(f)]$, comes from the integrated change in the orbital phase, 
caused by the precession of the orbital angular momentum vector that accompanies the spin precessions,
and is given by~\cite{Apostolatos:1994mx}
\begin{equation}
\delta_p \Phi[t(f)] = - \int_f^{f_{\rm final}} df \frac{2\mathbf{\hat{L}}\cdot
\mathbf{\hat{n}}}{1 - (\mathbf{\hat{L}}\cdot
\mathbf{\hat{n}})^2}(\mathbf{\hat{L}}\times \mathbf{\hat{n}})\cdot
\mathbf{\dot{\hat{L}}} \, .
\label{eq:dTomphase}
\end{equation}

In contrast to the case where the spins are aligned with the orbital angular momentum, precessions of the spins and of the orbital plane induce modulations of both the amplitude and phase of the gravitational wave on a precession time scale.   The orbital time scale is given by
\beq
T_{\rm orbital} \sim ( r^3 / M )^{1/2} \, ,
\eeq
while the precession time scale is given by
\beq
T_{\rm precession} 
\sim \frac{r^3}{\mathbf{L}} \sim \frac{r^{5/2}}{\mu M^{1/2}} \, .
\eeq
The final relevant time scale is that of the inspiral, given by
\beq 
T_{\rm inspiral}  \sim \frac{r^4}{\mu M^2} \, .
\eeq
Since for most of the inspiral, $T_{\rm orbital} \ll T_{\rm precession} \ll T_{\rm inspiral}$, we are justified to use orbit
averaged equations for the spin and angular momentum precessions, and to allow the total angular momentum $\bf J$ to evolve adiabatically as a result of gravitational radiation damping. 
 
The relevant equations~\cite{Apostolatos:1994mx} are
\bes
\begin{eqnarray}
    \mathbf{\dot{S}}_1 &=& \mathbf{\Omega}_1 \times \mathbf{S}_1 \, ,
    \label{eq:S1dot}\\
    \mathbf{\dot{S}}_2 &=& \mathbf{\Omega}_2 \times \mathbf{S}_2 \, ,
  \label{eq:S2dot} \\
     \mathbf{\dot{L}} &=& \mathbf{\dot{J}} - \mathbf{\dot{S}}_1 - \mathbf{\dot{S}}_2 
                           \, ,
  \label{eq:Ldot}
\end{eqnarray}
\ees 
where,  
\bes
\begin{eqnarray}
\mathbf{\Omega}_1 &=& \frac{1}{r^3} \, \left[
    \left(2+\frac{3}{2}\frac{m_2}{m_1}\right)\mu
    \sqrt{Mr} \mathbf{\hat{L}} 
\nonumber \right. \\
&& \quad 
\left .
-\frac{3}{2}(\mathbf{S}_2\cdot
    \mathbf{\hat{L}})\mathbf{\hat{L}} + \frac{1}{2} \mathbf{S}_2 \right] \, ,
    \label{Omega1} \\
\mathbf{\Omega}_2 &=&   \frac{1}{r^3} \, \left[ 
    \left(2+\frac{3}{2}\frac{m_1}{m_2}\right)\mu
    \sqrt{Mr} \mathbf{\hat{L}} 
\nonumber 
\right. \\
&& \quad
\left .
-\frac{3}{2}(\mathbf{S}_1\cdot
    \mathbf{\hat{L}})\mathbf{\hat{L}} + \frac{1}{2} \mathbf{S}_1  \right] \, ,
    \label{Omega2} 
\end{eqnarray}
\label{Omega12}
\ees
are the orbit-averaged precession vectors, and
\beq
\mathbf{\dot{J}} = - \frac{32}{5} \frac{\mu^2}{r} (\frac{M}{r})^{5/2} \mathbf{\hat{L}} \, ,
\label{eq:Jdot}
\eeq
is the change in the total angular momentum due to radiation reaction to lowest PN order and for a quasi-circular orbit.
 The overdot denotes a usual time derivative.

From Eqs. (\ref{eq:S1dot}) and (\ref{eq:S2dot}),  the magnitudes of the spin vectors $S_{i}$ do not change,
so the dimensionless spin parameters $\chi_i$ are constant.
 When $|\mathbf{L}| \gg |\mathbf{S}|$, spin-orbit coupling dominates, and  
the rate of precession of each spin is independent of the other spin.  In other words, when spin-orbit effects dominate, binaries with slowly spinning
objects produce roughly as many precession cycles as do binaries with faster spinning objects.
The difference is that for small $|\mathbf{S}|$ the cone describing the precession of $\mathbf{L}$ is smaller.  

In the general case of arbitrary initial conditions for the spin and 
angular momentum vectors the above system of equations cannot be solved 
analytically, so we must resort to numerical integration using routines 
from \cite{press}.

\section{Parameter estimation}
\label{Parameter}

We carry out the parameter estimation using the standard theory of Fisher information matrices 
and the maximum likelihood approximation developed for gravitational-wave applications by several authors
\cite{1992PhRvD..46.5236F,1994PhRvD..49.2658C,1995PhRvD..52..848P}.

Given the noise spectrum of the instrument and a signal $h(t;
\theta^a)$ characterized by a number of parameters
$\theta^a$ of the source, one can define the inner product between
two signals $h_1(t)$, $h_2(t)$ as follows, 
\bea
(h_1|h_2) 
&\equiv& 2
\int_0^{\infty} df \frac{ {\tilde{h}_1}^* (f) \tilde{h}_2(f) +
{\tilde{h}_2}^*(f) \tilde{h}_1 (f) }{S_n(f)}  
\nonumber \\
&=& 4 \; {\rm Re} \int_0^{\infty} df 
\frac{\tilde{h}_1^* (f)  \tilde{h}_2 (f) } {S_n(f)} \,, 
\label{Eq:innerproduct}
\eea 
where $\tilde{h}_1(f)$ and $\tilde{h}_2(f)$ are the Fourier
transforms of the respective gravitational waveforms
$h_i(t,\theta^a)$, and the star denotes complex conjugate, and $S_n(f)$ is the noise spectral density of the detector.  The
signal-to-noise ratio (SNR) for a given signal $h(t)$ is then given
by,
\begin{equation}
\rho[h] \equiv  (\tilde{h}|\tilde{h})^{1/2} \,, \label{Eq:SNR}
\end{equation}
evaluated at the estimated values $\theta^a$ of the source
parameters. In our analysis we will include the possibility that both detector combinations of {\em LISA}
will be operational. In this case, the Fisher information matrix $\Gamma_{ab}$ 
of the source is defined as follows,
\begin{equation}
\Gamma_{ab} \equiv \left( \frac{\partial h^{1}}{\partial\theta^a} \mid
\frac{\partial h^{1}}{\partial\theta^b} \right) + 
\left( \frac{\partial h^{2}}{\partial\theta^a} \mid
\frac{\partial h^{2}}{\partial\theta^b} \right) \,.
\label{Eq:FisherMatrix}
\end{equation}
where $h^{1}$, and $h^{2}$ are the signals in the two {\em LISA} arm combinations discussed earlier. 
In the limit of large SNR and if the noise is stationary and Gaussian, the
probability that the GW signal $h(t)$ is characterized by a given set of
values of the source parameters $\theta^a$ is
\beq
p(\mbox{\boldmath$\theta$}|h)=p^{(0)}(\mbox{\boldmath$\theta$})
\exp\left[-\frac{1}{2}\Gamma_{ab}\Delta \theta^a \Delta \theta^b\right]\,,
\eeq
where $\Delta \theta^a = \theta^a - {\hat \theta}^a$, is the difference
between the estimated and the true value of the parameters,
and $p^{(0)}(\mbox{\boldmath$\theta$})$ is the prior information.
An estimate of the rms error in measuring the
source parameter $\theta^a$ can then be calculated, in the limit of large
SNR, by taking the square root of the diagonal elements of the inverse
of the Fisher matrix,
\begin{equation}
\Delta\theta_{\rm rms}^a = \sqrt{\Sigma^{aa}} \,, \qquad  \Sigma = \Gamma^{-1} \,.
\label{errors}
\end{equation}
Finally, the correlation coefficients between two parameters $\theta^a$ and
$\theta^b$ are given by
\begin{equation}
c_{ab} = \frac{\Sigma^{ab}}{\sqrt{\Sigma^{aa}\Sigma^{bb}}} \,.
\label{Eq:correlations}
\end{equation}

It turns out that because {\em LISA} is designed
to detect massive inspirals, will naturally provide
the largest lower bounds on $\lambda_g$.
This can be seen from the dependence of the bound on $\lambda_g$ on the relevant parameters of the system and the detectors, given by Eq. (4.9) of \cite{1998PhRvD..57.2061W}:
\begin{equation}
\lambda_g \propto \left ( \frac{I(7)}{\Delta} \right )^{1/4}
\left ( \frac{D}{(1+Z)D_L} \right )^{1/2} \frac{{\cal M}^{11/12}}{S_0^{1/4} f_0^{1/3}} \,,
\label{lambdappnl}
\end{equation}
where $S_0$ is a parameter that establishes the floor of the noise spectral density (in Hz$^{-1}$), $f_0$ is a characteristic ``knee'' frequency, or frequency where the noise is a minimum.  The quantities $I(7)$ and $\Delta$ are determined from the Fisher matrix inversion and are largely independent of either $S_0$ or $f_0$, or of the SNR of the signal.  In any case, the bound is only weakly dependent on these variables.  The ratio $D/(1+Z)D_L$ is weakly dependent on distance, reflecting the fact that the effect of the massive graviton and the estimation errors both grow with distance.   Finally,  the factor $S_0^{1/4} f_0^{1/3}$ is roughly the same for {\em LISA} as it is for, say, advanced LIGO, and thus the best bound on $\lambda_g$ will come from {\em LISA}.

The noise spectrum of {\em LISA} consists of the instrumental noise intrinsic to the on-board instrumentation and drag-free control, and astrophysical noise due to
unresolved astrophysical sources of GW lying in the instrument's frequency band.
The instrumental noise currently used in the literature is that of reference
\cite{PhysRevD.62.062001} (also found online at \cite{LISAnoise}).
In our calculations we use an analytic version of the instrumental noise following 
\cite{2005PhRvD..71h4025B}, given by, 
\bea S_h^{\rm instr} (f) &=& 
\left[9.18\times 10^{-52}\,f^{-4}
+1.59\times 10^{-41}
\right. \nonumber \\
&& \left . 
+9.18\times 10^{-38}\, f^2\right]~{\rm Hz}^{-1}\,,
\label{eq:Instr.noise} 
\eea 
where $f$ is in Hz.  Technically this model ignores the oscillatory effects in the transfer function of LISA at high frequencies where the gravitational wavelength becomes comparable to the spacecraft separations, but since the relevant systems for bounding the graviton mass are massive binary inspirals at the low frequency end, we do not expect this simplification to have a large effect.

The spectral density for the noise from galactic binaries is approximated 
by \cite{2001A&A...375..890N},
\begin{equation}
S_h^{\rm gal} (f) = 2.1 \times 10^{-45} 
\,f^{-7/3} \; {\rm Hz}^{-1} \, ,
\label{eq:Sgal}
\end{equation}
and for extra-galactic binaries by \cite{2003MNRAS.346.1197F},
\begin{equation}
S_h^{\rm ex-gal} (f) = 4.2 \times 10^{-47} 
\, f^{-7/3} \; {\rm Hz}^{-1} \, .
\label{eq:Sexgal}
\end{equation}
The total noise spectrum to be used~\cite{2004PhRvD..70l2002B} is given by
\bea
S_h (f) &=& {\rm min}
\left \{  S_h^{\rm instr}(f) {\rm exp} \left( \kappa
T^{-1}_{\rm mission} dN/df \right), 
\right. \nonumber \\
&& \left .  S_h^{\rm instr}(f) + S_h^{\rm gal}(f)
\right \} + S_h^{\rm ex-gal} (f) \, ,
\label{Eq:tot.noise} 
\eea 
where $T_{\rm mission}$ is the duration of the mission, which we assume to be one year,
$\kappa = 4.5$ is the average number of frequency bins that are lost 
when each galactic binary is fitted out, and 
${dN}/{df} = 2 \times 10^{-3} {f}^{11/3} \,{\rm Hz^{-1}}$. 

In calculating the integrals for the Fisher matrix, we use the following expressions for the lower and upper limits 
of integration~\cite{2005PhRvD..71h4025B}.  The initial frequency is given by 
\bea
\label{fininital}
f_{\rm initial} &=&{\rm max} \left \{ f_{\rm low}, f(T_{\rm obs}) \right \} \,, 
\nonumber \\
 f(T_{\rm obs}) &=& 4.149\times 10^{-5}\left(\f{{\cal M}}{10^6 M_\odot}\right)^{-5/8}
\left(\f{T_{\rm obs}}{\rm yr}\right)^{-3/8}
 \, ,
\eea
where $f(T_{\rm obs})$ comes from the leading term of Eq. (\ref{eq:PNtime}), with $T_{\rm obs} = t_c - t(f)$, and where
$f_{\rm low}$ is the lower cutoff of the {\em LISA} instrument, taken 
here to be $10^{-5} \; {\rm Hz}$.  The final frequency is given by 
\beq
f_{\rm final} = {\rm min} \left\{ f_{\rm ISCO}, f_{\rm end} \right\}  \, ,
\eeq
where $f_{\rm ISCO} = (6^{3/2}\pi M)^{-1}$
is the usual frequency for the innermost stable circular orbit and 
$f_{\rm end}=1\; {\rm Hz}$ is a conventional upper cutoff for the {\em LISA} noise curve.   In order to see clearly the effects of spin-precessions on the graviton-mass bound, we choose the same observation time $T_{\rm obs} = 1$ year as in \cite{2005PhRvD..71h4025B}.
\section{Results}
\label{calculations}

In general a quasi-circular binary black hole inspiral in GR is described by a total of 15 parameters; adding the parameter for the massive graviton, we have the following 16 parameters:
the two individual masses of the system, $\ln(m_1)$, $\ln(m_2)$, 
the luminosity distance to the source $\ln(D_{L})$, 
the two dimensionless spin parameters $\chi_{\rm 1}$, $\chi_{\rm 2}$, 
the time and phase at coalescence $t_{c}$, $\Phi_{c}$,  
the two angles of the binary's sky position $\bar{\phi}_{S}$, $\cos\bar{\theta}_{S}$,
the two angles of the initial orientation of the orbital angular momentum vector, 
$\bar{\phi}_{L}$, $\cos \bar{\theta}_{L}$,
the four angles of the initial orientations of the spins of the two bodies, 
$\bar{\phi}_{S_1}$, $\cos \bar{\theta}_{S_1}$, $\bar{\phi}_{S_2}$, $\cos \bar{\theta}_{S_2}$,
and finally $\beta_{\rm g}$, the parameter that describes the  massive graviton contribution to the phase of
the waveform.
All angles are defined in the frame 
attached to solar system barycenter.

The inclusion of spin precessions makes some of
the parameters used traditionally for estimation less suitable.  For example, the spin-orbit 
and spin-spin parameters $\beta$ and $\sigma$ are now time (frequency) dependent, so one must either go directly to the values of $\chi_1$ and $\chi_2$ and the four initial spin orientation angles as parameters, or one must use the initial values of $\beta$ and $\sigma$ along with four other suitable parameters (such as the initial angles) as the appropriate parameters.  We choose the former.   Instead of
the chirp mass (${\cal M}$) and the symmetric mass ratio ($\eta$), we use the individual masses as parameters, because they more directly scale the spins.

Proceeding with the error estimation, we first fix a pair of masses in the source rest frame, 
the phase at coalescence and a redshift or luminosity distance.  We then randomly select 
the dimensionless spin parameters, $\chi_i$ within the range $[0,1]$, and the initial spin, orbital
angular momentum and source position angles (eight angles). We also select randomly the time
of the coalesence $t_{\rm c}$, within the assumed time duration of the mission, 
which corresponds to different orientations of the {\em LISA} 
arms at the first reception of the gravitational wave signal.
One effect of the selection of random $t_{\rm c}$, is that some signals might 
be partially cut off because they are already in the sensitive band when {\em LISA} starts observing them.  We use the routine RAN2~\cite{press} to produce the random numbers.

We also set the nominal value $\beta_{\rm g}=0$ for our calculations since we are interested in setting
a lower limit for $\lambda_{\rm g}$.

The inclusion of spin precessions modulates both the amplitude and the phase of the waveform. 
Since the total angular momentum ${\bf J} = {\bf S}_1 + {\bf S}_2 + {\bf L}$  
is conserved on a precession timescale, the orbital angular momentum vector ${\bf L}$ must precess to cancel 
out the effects of spin precessions. 
As a consequence, the amplitude given by expression (\ref{eq:polamp}), 
now changes and modulates the waveform accordingly. 
The phase is also affected mainly through the 
terms that describe the polarization phase (\ref{eq:polphase}) 
and integrated change in orbital phase (\ref{eq:dTomphase}). Finally, in the phasing
function $\Psi(f)$ the parameters $\beta$ and $\sigma$ 
are now frequency dependent.

Another thing to note about precession is the following. Since we generate 
arbitrarily the initial directions of the spin and angular momentum vectors we can have
both kinds of precession, simple and transitional, as described in 
\cite{Apostolatos:1994mx}. ``Simple'' precession is the (most common) case where the angular momentum vector
$\mathbf{L}$ and the total spin vector $\mathbf{S}$ precess around the total angular
momentum vector $\mathbf{J}$, which decreases slowly because of gravitational radiation reaction.  Simple precession always occurs when $|\mathbf{L}| \gg |\mathbf{S}|$, which is generally the case early in the inspiral.  
``Transitional'' precession occurs when $\mathbf{{L}} $ 
and $\mathbf{{S}}$  are almost antialigned and $|\mathbf{L}| < |\mathbf{S}|$.   It consists of a ``tumbling'' of the
$\mathbf{{L}}$ and $\mathbf{{S}}$ vectors (with the sum still tied to $\mathbf{{J}}$) because of the
loss of ``gyroscopic bearings'' of the system. 

Apostolatos {\em et al.}~\cite{Apostolatos:1994mx} found that, in order to get transitional precession, the initial angle between the total 
spin ${\bf S}$ 
and the angular momentum ${\bf {L}}$ must be larger than about $164^{\circ}$, so that, as $|{\bf {L}}|$ decreases because of radiation reaction, the conditions for transitional precession will be met during the inspiral phase.
We have checked our initial values and learned that out of the $10^4$ sets of initial 
angles for ${\bf \hat{S}_1}$, ${\bf \hat{S}_2}$ and ${\bf \hat{L}}$ only about 80 lead to transitional precession, and for these, the Fisher matrix calculations were not adversely affected by the complicated precessions.

Our calculations start with the numerical integration of the spin precession equations
(\ref{eq:S1dot})-(\ref{eq:Ldot}) in the frequency domain, using the random initial values
for the six parameters of the spins of the two bodies and the two components of the orbital
angular momentum, to get the orientations of ${\bf \hat{L}}$, ${\bf \hat{S}_1}$, ${\bf \hat{S}_2}$ 
over the duration of the signal.  We use Eq.\ (\ref{eq:fdot}) to convert from $d/dt$ to $d/df$, and use Kepler's third law at lowest order, $r = M^{1/3}/(\pi f)^{2/3}$ to convert from $r$ to $f$ in Eq.\ (\ref{Omega12}).   We use a fourth order Runge-Kutta constant step size 
routine RK4~\cite{press}.  Once this is done, the spin parameters $\beta$ (spin-orbit), 
$\sigma$ (spin-spin) and the integrated phase correction
$\delta_p \Phi[t(f)]$  (\ref{eq:dTomphase}) 
are calculated.  Subsequently the signal in the frequency domain, Eq.\ (\ref{eq:freqdomainsignal}), 
is calculated on the same grid on which the precession equations are solved.

\begin{figure}[!t]
\begin{center}
\includegraphics[width=2.8in,angle=-90]{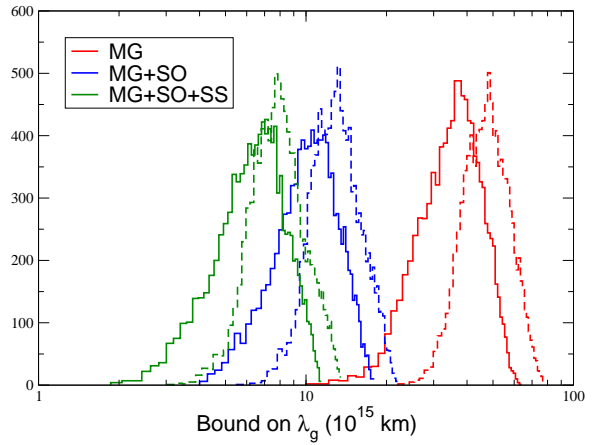}
\caption{Distribution of lower bounds on the graviton Compton wavelength 
$\lambda_{g}$ (in units of $10^{15}$ km ) for $10^4$ binaries when spin is included without precession. The system is a $10^6$ + $10^6$ $M_{\odot}$ BBH at $z= 0.55$ (3 Gpc). 
Solid (dashed) lines refer to one (two) {\em LISA} detectors. Number of bins is set to 50. Results match Fig.\ 6 of~\cite{2005PhRvD..71h4025B} }
\label{fig:betaGnospin}
\end{center}
\end{figure}

\begin{figure}[!t]
\begin{center}
\includegraphics[width=2.8 in,angle=-90]{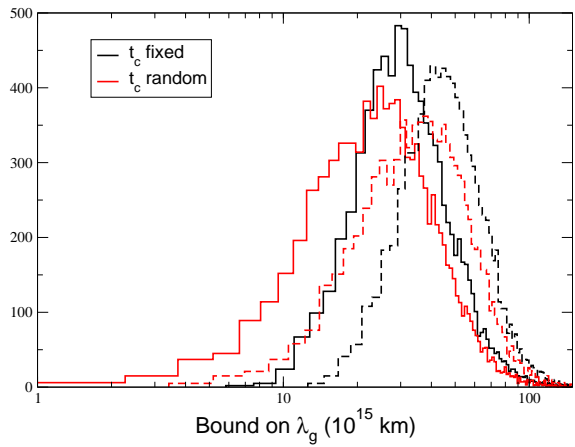}
\caption{Distribution of lower bounds on $\lambda_{g}$ (in units of $10^{15}$ km)
for $10^4$ binaries including spin precessions.  The system is a $10^6$ + $10^6$ $M_{\odot}$ BBH at $z= 0.55$ (3 Gpc).
Red curve is for $t_{c}$ fixed to one year; blue curve is for
random values of $t_{c}$ in the one year interval of the {\em LISA} mission.
Solid (dashed) lines refer to one (two) {\em LISA} detectors respectively. }
\label{fig:Tc_compare}
\end{center}
\end{figure}

All the derivatives of the signal with respect to parameters needed for the Fisher matrices 
are calculated numerically.  Given a determination of $h(f)$ for a given set of initial values of the 16 parameters $\theta^a$, we also calculate $h(f)$ for  nearby values $\theta^a + \delta \theta^a$ and 
$\theta^a - \delta \theta^a$ for each parameter in turn.  Then for each $\theta^a$, we calculate the derivative  using the standard central finite difference formula,
\beq
\frac{\partial h(f)}{\partial \theta^a} \simeq \frac{h(f; \theta^a + \delta \theta^a)
- h(f;\theta^a - \delta \theta^a)}{2 \; \delta \theta^a} + {\cal O}(\delta {\theta^a}^2 ) \, ,
\eeq
for each value of $f$ on the grid.
Since we are using double precision accuracy for our variables a natural choice of
the small shifting parameter $\delta\theta^a$ for the calculation of the numerical derivatives
would be $\delta\theta^a \simeq 10^{-7} - 10^{-8} $.  We have chosen $\delta\theta^a = 10^{-8}$ 
for all the sixteen parameters estimated in order to achieve the best possible accuracy.
Then the necessary integrals are calculated numerically on the same grid 
using the extended Simpson rule for the closed interval $[f_{\rm initial},f_{\rm final}]$. 

\begin{figure}[t]
\begin{center}
\includegraphics[width=2.8in,angle=-90]{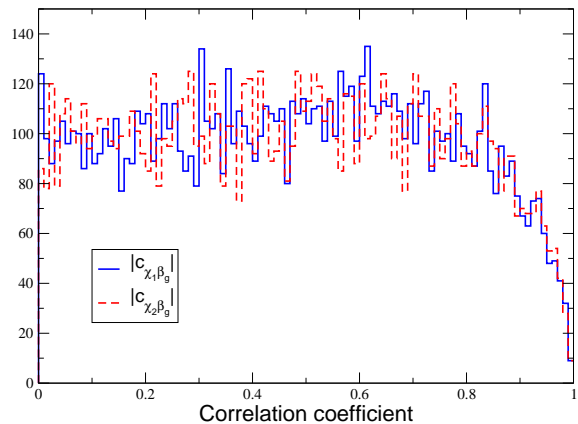}
\includegraphics[width=2.8in,angle=-90]{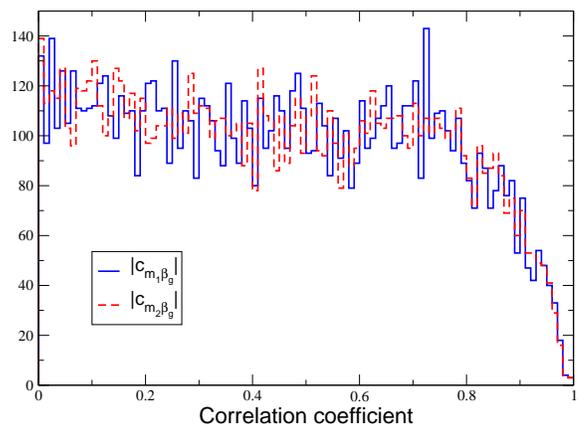}
\caption{Distribution of the correlation coefficients between spin parameters $\chi_i$ 
and $\beta_{g}$ and  
between individual masses $m_i$ and $\beta_{g}$}
\label{fig:corrcoeff}
\end{center}
\end{figure}

Finally, the inverse of the Fisher matrix is calculated using the routine SVDCMP~\cite{press}.
We have also used LU and Gauss-Jordan decomposition as a cross check, with identical results. 
The main advantage of this routine is that it allows us to check whether the matrix is 
ill-conditioned for the inversion,
by calculating the ratio of the smallest to the largest eigenvalue of the matrix. 
If this ratio is of the order of the machine accuracy  ($\simeq 10^{-16}$), then the matrix inversion is not to be trusted. 
However, another simple sanity test is to
multiply the original with the produced inverse matrix 
and check how far  the product is from the identity matrix. 
This can be measured by the maximum deviation of the non-diagonal elements of the matrix 
from zero ($\simeq 10^{-16}$) in double precision.  We have checked both of these criteria. 
In most of the cases the condition number is of the order of $10^{-10}-10^{-11}$ and
the maximum deviation is of the order of $10^{-6}-10^{-7}$.
In the cases where the condition number approaches double precision accuracy i.e. 
$\simeq 10^{-16}$, the maximum deviation is of the order of $10^{-4}$.

We have carried out several tests and diagnostics for the validity of our code.
In the case of aligned, nonprecessing spins, we have reproduced the fifth panel of Fig.\ 6 of Berti {\em et al.}~\cite{2005PhRvD..71h4025B} for one year integration time and $t_{\rm c}$ fixed; the results, shown in Fig.\ \ref{fig:betaGnospin}, agree within the natural statistics of our Monte Carlo simulations.   
In contrast to \cite{2005PhRvD..71h4025B}, we have been able to quote errors on the graviton mass including spin-spin effects
because of the improved machine precision available that allowed us to invert the larger matrices reliably.
Also for the non-precessing case, and for individual choices of angles, we have compared parameter estimation errors and correlation 
coefficients with those from a Mathematica code developed independently and used by K. G. Arun for other
calculations; the agreement was excellent. 

For the precessing cases, we have checked our code with respect to 
the median error results quoted by Lang and Hughes \cite{2006PhRvD..74l2001L,2008ApJ...677.1184L} for the asymmetric mass 
systems of $(m_1,m_2) = (3,1) \times 10^5 M_{\odot}$ and 
$(m_1,m_2) = (3,1) \times 10^6 M_{\odot}$ at $z=1,3,5$ respectively. 
Modulo the statistics of the Monte Carlo simulation, we  found good agreement for the median errors in masses and dimensionless spin parameters, 
the semi-major and semi-minor
axis values $(a,b)$ of the error ellipse on the sky, and the luminosity distance
and angular resolution; the comparisons are shown in
tables \ref{table:comparez1}-\ref{table:comparez5}.

The effect of choosing arbitrary coalescence times $t_c$ is illustrated in Fig.\ \ref{fig:Tc_compare} where the distributions of lower bounds
on the graviton Compton wavelength $\lambda_{g}$ are shown for 
fixed and random values of $t_{c}$ . It is clear from the graph that randomizing $t_{c}$, leads to somewhat smaller lower
bounds, with a tail at low values of the bounds, depicting the effect of signal loss in some of the cases.   Fig.\ \ref{fig:Tc_compare} also shows that using two {\em LISA} arm combinations generally leads to improved bounds.  

In Fig.\ \ref{fig:corrcoeff} we plot the distribution of the correlation coefficients between the massive graviton 
parameter $\beta_{\rm g}$, and the two dimensionless spin parameters $\chi_i$ (top panel) and the 
two masses $m_i$ (bottom panel) for the $10^6 + 10^6 \, M_\odot$ black hole case.  The correlations are quite mild, with most of the 
values ranging between 0 and 0.8, in contrast to
the nonprecessing case \cite{2005PhRvD..71h4025B}, where correlation coefficients larger than $0.9$ were routine.  This illustrates the strong decorrelating effect of the precessions.

\section{Conclusions} 
\label{conclusions}

In this paper we have studied bounds that can be placed on the mass of a hypothetical graviton using GW observations from the planned {\em LISA} mission, including spin precession
effects.  
One possible extension of this work would be to include the effect of 
higher amplitude harmonics of the GW signal; in the non-spinning case, this is known to improve the accuracy of estimating parameters, including distance and sky location (see, eg. \cite{2007PhRvD..76j4016A, 2008PhRvD..77b4030T,2008PhRvD..78f4005P}), and the graviton mass \cite{Arun:2009pq}. 
A final note: The inclusion of spin precessions
has a significant computational cost in the parameter estimation procedure. 
Recently Kocsis {\em et al.}~\cite{Kocsis:2007hq} have developed a very efficient way, 
the harmonic mode decomposition, of decoupling the several parameters 
that the signal depend on, according to their frequency ``signature''.
This way the integrations for computing the elements of
the Fisher matrices can all be done at once, lowering significantly the computational cost. 
It would be interesting to try to implement this decomposition on our code in the
future.

\section{Acknowledgments}
It is a pleasure to thank K.G. Arun for many useful conversations and suggestions 
during the last stages of this work as well as for help with the calibration of the 
code in the non precessing limit. 
A.S. also benefited from private communications with Emanuele Berti and Ryan N. Lang.   The anonymous referee made several useful suggestions for improvement.
This work was supported in part by the National Science Foundation Grant No. 
PHY 06-52448, the National Aeronautics and Space Administration, Grant No. 
NNG-06GI60G, and the Centre National de la Recherche Scientifique, Programme Internationale de la Coop\'eration Scientifique (CNRS-PICS), Grant No. 4396.   We are grateful for the hospitality of the Institut d'Astrophysique de Paris, where parts of this work were carried out.  All the computer runs for this paper were performed at the HPC center 
of the Physics Department of Washington University.

\bibliography{references}


\begin{widetext}

\begin{table}[!b]
\begin{center}
\begin{tabular}{|c|c||c|c|c|c||c|c||c|c|}
\hline
$m_1\ (M_\odot)$ & $m_2\ (M_\odot)$ & $\Delta m_1/m_1$ & $\Delta m_2/m_2$ & $\Delta \chi_1$ & $\Delta \chi_2$ & $2 a$ (arcmin) & $2 b$ (arcmin) & $\Delta \Omega_S$ $\rm {(deg^2)}$ & $\Delta D_L/D_L$ \\
\hline
$3\times 10^5$ & $10^5$ & 0.000667 & 0.000541 & 0.00157 & 0.00306 & 16.9  & 7.3 & 0.0233 & 0.00240  \\ 
& & {\it 0.000387}& {\it 0.000314} & {\it 0.00130} & {\it 0.00176}& {\it 13.9} & {\it 8.4} & {\it 0.0245} & {\it 0.00236} \\
$3\times 10^6$ & $10^6$ & 0.00238 & 0.00192 & 0.00380 & 0.00674 & 32.3 & 14.7 & 0.0839 & 0.00419 \\ 
& & {\it 0.00458}& {\it 0.00371} & {\it 0.00357} & {\it 0.00613}& {\it 23.8} & {\it 14.6} & {\it 0.0730} & {\it 0.00193} \\
\hline
\end{tabular}
\caption{Comparison of median errors in selected parameters for two cases of $10^4$ precessing binaries
 at $z = 1$.  Semi-major axes of error ellipse on the sky parametrized by $a$ and $b$; angular resolution is $\Delta \Omega_S$.  Lang-Hughes results are quoted in the first line; our
results are in italics in the second line.}
\label{table:comparez1}
\end{center}
\end{table}

\begin{table}[!b]
\begin{center}
\begin{tabular}{|c|c||c|c|c|c||c|c||c|c|}
\hline
$m_1\ (M_\odot)$ & $m_2\ (M_\odot)$ & $\Delta m_1/m_1$ & $\Delta m_2/m_2$ & $\Delta \chi_1$ & $\Delta \chi_2$ & $2 a$ (arcmin) & $2 b$ (arcmin) & $\Delta \Omega_S$ $\rm {(deg^2)}$ & $\Delta D_L/D_L$ \\
\hline
$3\times 10^5$ & $10^5$ & 0.00363 & 0.00294 & 0.00879 & 0.0171 & 92.5 & 32.5 & 0.656 & 0.0126 \\ 
& & {\it 0.00225}& {\it 0.00182} & {\it 0.00671} & {\it 0.0140}& {\it 83.5} & {\it 49.0} & {\it 0.885} & {\it 0.0058} \\
$3\times 10^6$ & $10^6$ & 0.0181 & 0.0148 & 0.0223 & 0.0386 & 142 & 64.6 & 1.65 & 0.0193 \\ 
& & {\it 0.0129}& {\it 0.0103} & {\it 0.0130} & {\it 0.0290}& {\it 96.8} & {\it 58.4} & {\it 1.21} & {\it 0.0161} \\
\hline
\end{tabular}
\caption{Same as Table \ref{table:comparez1}, but for $z=3$}
\label{table:comparez3}
\end{center}
\end{table}

\begin{table}[!b]
\begin{center}
\begin{tabular}{|c|c||c|c|c|c||c|c||c|c|}
\hline
$m_1\ (M_\odot)$ & $m_2\ (M_\odot)$ & $\Delta m_1/m_1$ & $\Delta m_2/m_2$ & $\Delta \chi_1$ & $\Delta \chi_2$ & $2 a$ (arcmin) & $2 b$ (arcmin) & $\Delta \Omega_S$ $\rm {(deg^2)}$ & $\Delta D_L/D_L$ \\
\hline
$3\times 10^5$ & $10^5$ & 0.00811 & 0.00658 & 0.0193 & 0.0359 & 217 & 95.8  & 3.73 & 0.0284 \\
& & {\it 0.00476}& {\it 0.00410} & {\it 0.0108} & {\it 0.0150}& {\it 201} & {\it 123} & {\it 5.22} & {\it 0.0145} \\
$3\times 10^6$ & $10^6$ & 0.0576 & 0.0475 & 0.0606 & 0.107 & 304 & 139 & 7.52 & 0.0436 \\ 
& & {\it 0.0656}& {\it 0.0536} & {\it 0.0468} & {\it 0.112}& {\it 190} & {\it 116} & {\it 4.68} & {\it 0.0164} \\
\hline
\end{tabular}
\caption{Same as Table \ref{table:comparez1}, but for $z=5$}
\label{table:comparez5}
\end{center}
\end{table}

\end{widetext}
\end{document}